\documentclass[amsmath,amssymb,aps,twocolumn,superscriptaddress,longbibliography]{revtex4-1}

\usepackage{braket}
\usepackage{mathtools}

\usepackage[dvipsnames]{xcolor}

\usepackage[colorlinks=false]{hyperref}

\usepackage{xargs}                      

\def\bea{\begin{eqnarray}}
\def\eea{\end{eqnarray}}
\def\ben{\begin{equation}}
\def\een{\end{equation}}
\def\benu{\begin{enumerate}}
\def\enu{\end{enumerate}}
\def\capt{\caption}

\usepackage{graphicx}
\usepackage{dcolumn}

\begin{document}

\preprint{APS/123-QED}

\title{Multi-sliced Gausslet Basis Sets for Electronic Structure}

\author{Steven R.\ White}
\affiliation{Department of Physics and Astronomy, University of California, Irvine, CA 92697-4575 USA}

\author{E.\ Miles Stoudenmire}
\affiliation{Center for Computational Quantum Physics, Flatiron Institute, New York, NY 10010 USA}

\date{\today}

\begin{abstract}
We introduce highly local basis sets for electronic structure which are very
efficient for correlation calculations near the complete basis set limit.
Our approach is based on gausslets, recently introduced wavelet-like smooth
orthogonal functions.  We adapt the
gausslets to particular systems using one dimensional coordinate
transformations, putting more basis functions near nuclei, while
maintaining orthogonality.  Three dimensional basis functions are composed
out of products of the 1D functions in an efficient way called
multislicing.  We demonstrate the new bases with both Hartree Fock and density 
matrix renormalization group (DMRG) calculations on hydrogen chain systems.
With both methods, we can go to higher accuracy in the complete basis set limit
than is practical for conventional Gaussian basis sets, with errors near 0.1 mH per atom.
\end{abstract}

\maketitle

Electronic structure calculations on molecules, solids, and biological systems
are performed by thousands of groups worldwide and account for a substantial
fraction of the world's scientific computing. Strongly correlated systems,
for which density functional approaches are inadequate, make up a small but
important fraction of these calculations.
An almost universal problem with methods for
strong correlation is poor computational scaling in both
system size and accuracy.  For example, wavefunction
methods, such as coupled cluster or configuration interaction,
typically scale as $N_e^6$ or higher for $N_e$ electrons.
Since correlation methods must deal with the two-electron interaction directly,
scaling of at least $N^4$ when using $N$ basis functions can appear hard to avoid,
since the two-electron interaction terms are described by a tensor $V_{ijkl}$ 
with four indices running over $N$ values.

A natural way to reduce the size of the $V_{ijkl}$ tensor is through local
basis functions.  In the extreme case of a grid descretization, the interaction
is reduced by a factor of $N^2$ to a matrix $V_{ij}$, with $V_{ij} = 1/|\vec r_i-\vec r_j|$
for $i \ne j$. For the less extreme case of basis sets where the functions have
substantial spatial compactness, we say that two basis functions $b_i$ and
$b_j$ ``overlap'' if there is some point $\vec r$ where $b_i(\vec r) b_j(\vec r)$
is significantly different from zero.  Terms in $V_{ijkl}$ are negligible
unless basis functions $i$ and $j$ overlap and also $k$ and $l$ overlap.  However, in three
dimensions, even for substantially localized functions, many basis functions
overlap, particularly if the functions have been orthogonalized, and one may not
realize a significant increase in the sparseness of $V_{ijkl}$ for systems
small enough to study feasibly.  This is unfortunate, since basis methods have
several advantages over grids, such as the ability to add extra atom-centered
core functions to better resolve the nuclear cusps.

Recently one of us introduced a novel basis function approach that has the same
favorable scaling of the interaction as a pure grid \cite{White:2017}.  This involved two key
ingredients: first, the introduction of a wavelet-related set of highly
localized, smooth orthogonal basis functions, called \emph{gausslets}, where
each function is defined as a sum over an underlying grid of simple
Gaussians.  Second, it was shown that one can construct an
accurate purely diagonal interaction $V_{ij}$ for a gausslet basis.  This
diagonal interaction for a special type of basis is not new in itself: a basis
of sinc functions also allows this construction \cite{Jones:2016}.  
 However, the extreme delocalization of sinc
functions is a severe disadvantage as a basis; the gausslet development in Ref.~\onlinecite{White:2017} 
shows that one can get the diagonal property with much more localized functions,
where it is based on the ability of gausslets to integrate like a delta
function. But the usefulness of gausslets was previously only demonstrated for 1D toy
systems.

Here we generalize the gausslet approach to three dimensions and practical
electronic structure calculations. Given a 1D basis, one can always
generate a 3D basis as coordinate products, i.e. $G_{ijk}(x,y,z) = f_i(x) g_j(y) h_k(z)$. 
This simple approach produces overly large basis sets. Instead, we introduce
coordinate transformations which put more functions near nuclei, and a procedure
called multislicing which allows the use of 1D coordinate transformations rather
than more complicated 3D transformations. Our multisliced gausslet (MSG) approach is a generalization of
our earlier sliced basis approach \cite{Stoudenmire:2017}.  We demonstrate the resulting method
on hydrogen chain systems \cite{Motta:2017}, using both Hartree Fock and the 
density matrix renormalization group (DMRG) \cite{White:1992,White:1999,Chan:2011}. 
In both cases the diagonal property allows for dramatically increased basis set size and high accuracy.
The combination of MSG and DMRG (MSG-DMRG) allows for simultaneously exact correlation and the complete
basis limit, going well beyond chemical accuracy in a controlled way. The MSG-DMRG approach
also features approximately linear scaling in one of the directions (which we take to be $z$) along which
the system is most extended.

\begin{figure}[t]
\includegraphics[width=0.9\columnwidth]{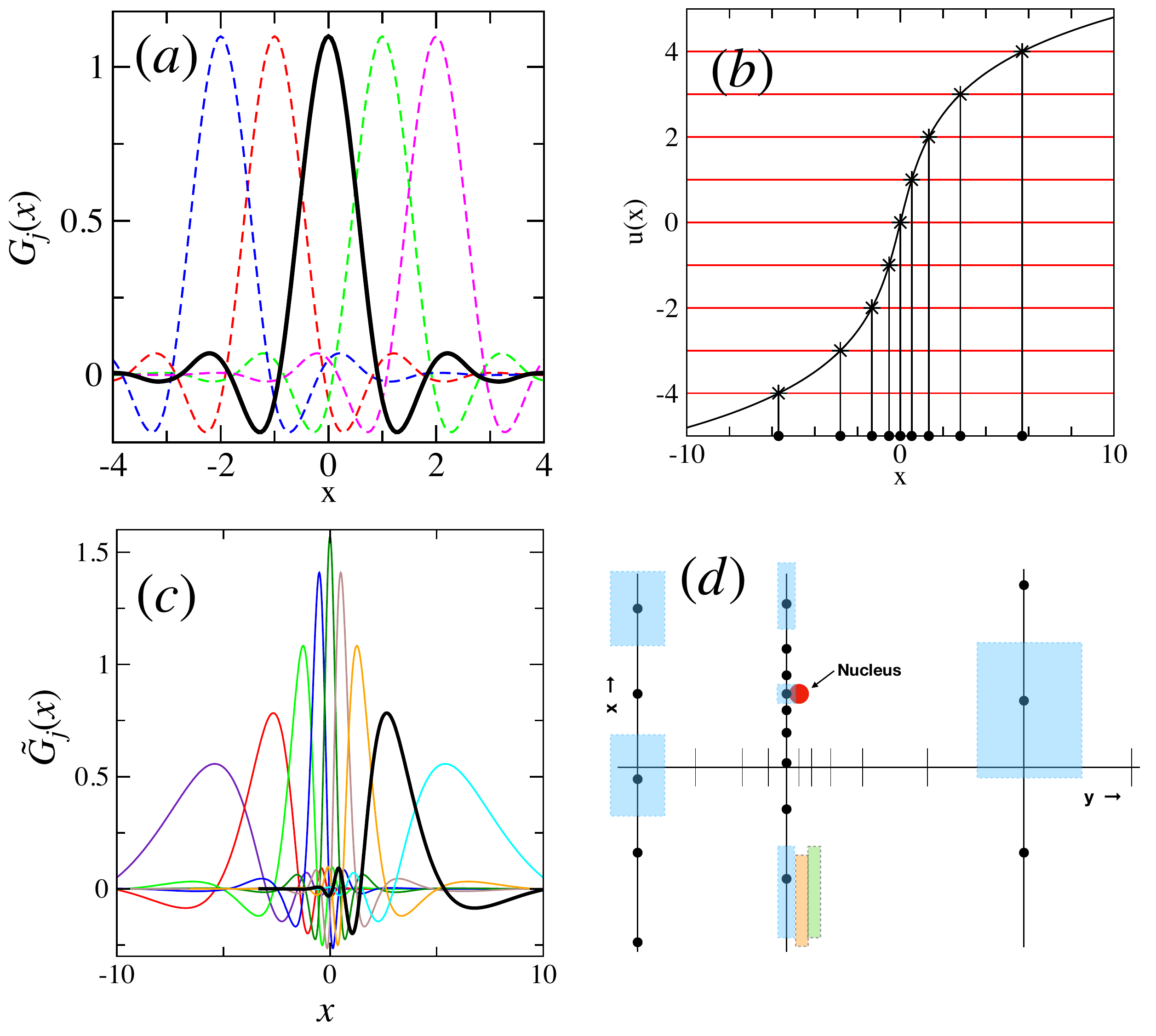}
\capt{(a) Array of gausslets, with the gausslet centered at the origin emphasized to show detail. 
(b) Coordinate transformation function $u(x)$ for a single atom, with $a=s=0.7$ in Eq. (\ref{oneatomdist}),
to give gausslets variable resolution.   
(c) Distorted gausslet basis based on the transformation of (b), which is orthonormal and 
allows a diagonal approximation. One of the functions is emphasized. (d) Schematic representation of multislicing
in 2D. The vertical lines represent slices, with three shown in detail. Each dot is the center of a basis function,
and the shaded rectangles illustrate the principle support region of some of the functions, although they have smooth tails
well beyond the rectangles.  The multicolored shaded rectangles represent long, thin basis functions which one would want
to contract at a later stage.
} \label{figabcd}
\end{figure}

In standard orthogonal wavelet theory, basis sets are made of two types of
functions, scaling functions, which carry low momentum, and wavelets, which carry
a range of higher momenta.  Gausslets are scaling-function-like 
functions, which are constructed out of sums of Gaussians for convenience.
A set of gausslets are shown in Fig.~\ref{figabcd}(a), highlighting a single gausslet 
in the center of the figure.
To make a 1D basis, one puts a
gausslet at each point on an evenly spaced grid, scaling them to match the grid spacing. 
The oscillatory tails make the gausslets precisely orthonormal, and
they can exactly represent polynomials up to some predetermined order (e.g. 10th order).

Modifying the gausslets with coordinate transformations allows spatially varying resolution.
Let $x(u)$ and its inverse $u(x)$ define a 1D smooth one-to-one coordinate mapping, which
will be used to make the grid narrow and closely spaced near nuclei, and wide and sparse far
away. First consider a 1D arrangement, with just one atom at $x=0$.
Define the gausslets on a uniform grid in
the $u$ space and then map to $x$-space, inserting a Jacobian factor to preserve
orthonormality.  If $G_j(u)$ is a gausslet centered
at integer $j$, define
\ben
\tilde G_j(x) = G_j(u(x)) \sqrt{u'(x)} \ .
\een
It is easy to see that the $\tilde G_j$ are orthonormal if the $G_j$ are.
An approximate local density of gausslets (i.e. inverse spacing) is $\rho(x) = u'(x)$.
To choose $u(x)$, consider moving from a nucleus with a high density
of basis functions to the tail region with a low density.  If one changes the density
too quickly, the ability of the basis to represent low-momentum functions will be
compromised.  It is natural to require the fractional change in $\rho$ when moving from
one gausslet to the next to be roughly constant. Thus, one wants
\ben
\frac{d\rho}{du} \propto \rho \ \, .
\een
This implies that $\rho$ falls off as $\sim 1/x$.  For small $x$, we need to choose
a maximum finite resolution, while keeping $\rho$ smooth.  For a single atom,
we choose
\ben
\rho = \frac{1}{s \sqrt{x^2+a^2}} \label{oneatomdist}
\een
where the parameter $s$, the {\it scale}, sets  or adjusts the overall gausslet spacing,
and $a$, the {\it core cutoff},  sets the range in $x$ over which we stop
decreasing the gausslet spacing.  The smallest gausslet spacing at the nucleus is about
$a\cdot s$.  This form for $\rho$ integrates to give 
\ben
u(x) = \sinh^{-1}(x/a)/s
\een
This transformation is shown in  Fig.~\ref{figabcd}(b), with the resulting 1D functions shown
in Fig.~\ref{figabcd}(c).

For multiple atom 1D systems, one can attempt to add a density of this form for each atom, but the long tails build up too high
a density in the center.  
In the Supplementary Material we describe the simple modification of summing the densities that we use
for many atoms 
\footnote{See Supplemental Material for more information about computing integrals for a multi-sliced
basis and contracting functions together using a technique based on Hartree-Fock. The Supplemental
Material also includes the additional references 
Refs.~\cite{Lin:2012,Babbush:2018,Julia}.}

If there were no coordinate transformation, we could make 3D basis functions 
as single products of gausslets $G_i(x) G_j(y) G_k(z)$.
This coordinate-product form greatly simplifies evaluation of integrals defining the Hamiltonian.  
To keep this
form, we apply coordinate transformations to each coordinate separately, in a method we call
``multislicing''.
In multislicing, the coordinate directions are sliced up sequentially, $z$, then $y$, then $x$.
A first coordinate transformation $u^z(z)$ determines a set
of $z$-values $z_k$ ($k=1,2,\ldots$), with $u^z(z_k) = k$, at which are centered distorted 1D gausslets $\tilde G_k(z)$.
The plane $z=z_k$ and function $\tilde G_k(z)$ together define a $z$-slice.
Next we slice up each $z$-slice in the $y$ direction, with a coordinate transformation unique to $k$,
$u^y_k(y)$, which defines a set of $y$-values $y_{kj}$. 
A $y$ ``slice'' (or ``subslice'' of a ``parent'' $z$-slice) is the line $z=z_k$, $y=y_{kj}$, 
with associated 2D function $\tilde G_k(z) \tilde G_{kj}(y)$.  
Finally, for each $y$-slice, define a unique coordinate transformation $u^x_{kj}(x)$, 
determining a set of $x$ values $x_{kji}$, and 3D basis functions 
$\tilde G_k(z)\tilde G_{kj}(y)\tilde G_{kji}(x)$.

The key point in using this successive procedure is to use of the knowledge of where a
slice is, relative to the nuclei, to make subsequent transformations with the lowest density of functions.
This is illustrated schematically in 2D in Fig.~\ref{figabcd}(d). Preserving the product form via multislicing
means that some basis functions are long and thin; however, at a later stage on can devise methods to contract
such functions with their neighbors, reducing unnecessary degrees of freedom. The details of the coordinate
transformations in the multisliced case are discussed in the Supplementary Material.

Each basis function has a well defined center $(x_{kji},y_{kj},z_k)$, and we can make a simple rule for
which functions to keep:  if the basis function is within a distance $b$ of an atom, we keep it.
Here $b=9$ a.u. proved very accurate ($<0.1$ mH errors compared to larger $b$)
except for $R=1$ for H$_{10}$, where we used $b=13$.

\begin{figure}[t]
\includegraphics[width=1.0\columnwidth]{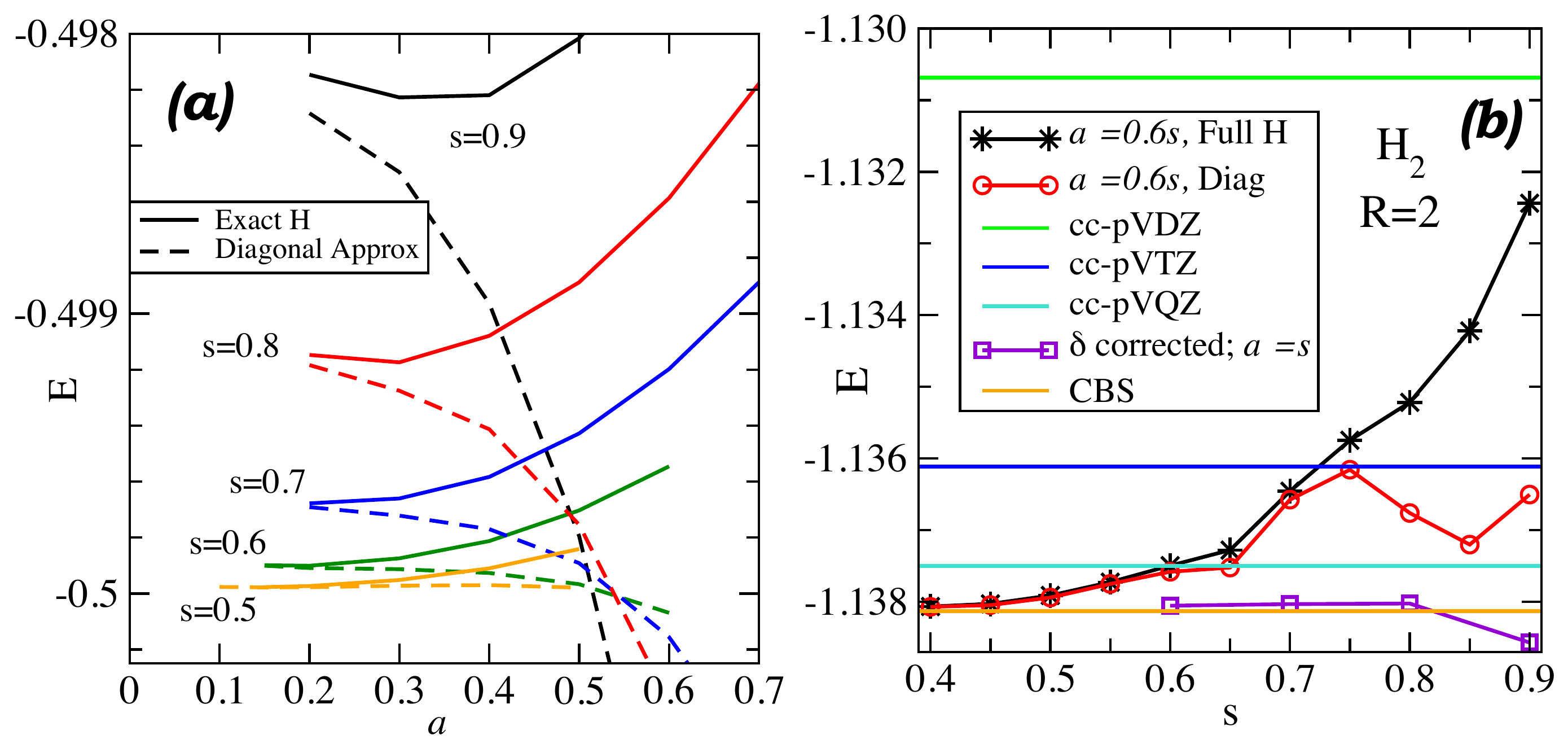}
\capt{(a)Energies of a hydrogen atom in an MSG basis as a function of $a$ and $s$, in Hartrees, using
the full Hamiltonian (``exact H") and using the integral diagonal approximation. 
The exact energy is $-1/2$.
(b) Energy of a hydrogen molecule with separation $R=2$ in standard Gaussian and MSG bases.
} \label{H1H2}
\end{figure}

Figure \ref{H1H2}(a) shows energies for a single hydrogen atom for various $a$ and $s$, using  
both the standard full Hamiltonian
and using a Hamiltonian matrix for which a diagonal approximation is made for the single particle 
potential \cite{White:2017}. Since there are only $N^2$ single particle terms, not $N^4$, using
this diagonal approximation barely improves computational efficiency, but one would expect this
approximation to mimic 
the performance of the more important two-particle diagonal approximation.
The diagonal approximation is sensitive to the singularity in the  potential at the nucleus, but 
increasing the basis function density near
the nucleus by decreasing $a$, for fixed $s$, nearly eliminates the diagonal approximation error.
A simple procedure to systematically converge to the ground state for this system would be to 
fix $a/s$ to be a constant, say 0.5-0.6, and then decrease $s$.


Figure \ref{H1H2}(b) shows the energy for a hydrogen molecule, 
compared to standard basis sets cc-pVxZ, where x=D, T, and Q,
and also compared to the exact energy from a treatment in special coordinates \cite{Kolos:1986}. 
A diagonal approximation for the two particle interaction is used here and in all 
subsequent MSG bases, since calculations would not be practical with the 
standard $V_{ijkl}$ form.
All results shown are exact (full CI) given the approximate Hamiltonian.
The MSG bases systematically converge to the exact results, and the diagonal
approximation for the single particle potential closely approximates the full Hamiltonian, 
particularly for smaller $s$.

Also shown in Fig. \ref{H1H2}(b) is a basis with a special delta-function correction for the nuclear cusp.
Increasing the resolution near nuclei by using a small $a$ is inefficient, leading to many basis functions. 
For example, for the hydrogen atom of Fig. \ref{H1H2}(a), taking $a=0.3$, $s=0.6$ produced 1179 functions, which resulted in an error of 0.13~mH. 
Our correction consists of adding a single-particle potential at each atom $\alpha$ of the form 
$v_\alpha \delta(\vec r-\vec r_\alpha)$.  
The parameter $v_\alpha$ is set by ``turning off" all nuclear electron potentials for 
atoms other than $\alpha$ (yet keeping the same set of functions to be used for the entire system),
and adjusting $v_\alpha$ so that the one-electron ground state energy is the exact 
hydrogen atom energy $-1/2$.
The errors associated with choosing $a$ too large are localized near the nuclei;  
the delta function potential alters
the terms in the Hamiltonian only for the basis functions overlapping with a nucleus.
Most importantly, $v_\alpha \to 0$ as $a\to0$ or $s\to 0$, so this correction does not change what
the basis converges to, only how fast it converges, accelerating the convergence.
In Fig. \ref{H1H2} and for the rest of the results, we set $a=s$ and use the delta correction.

\begin{figure}[t]
\includegraphics[width=0.8\columnwidth]{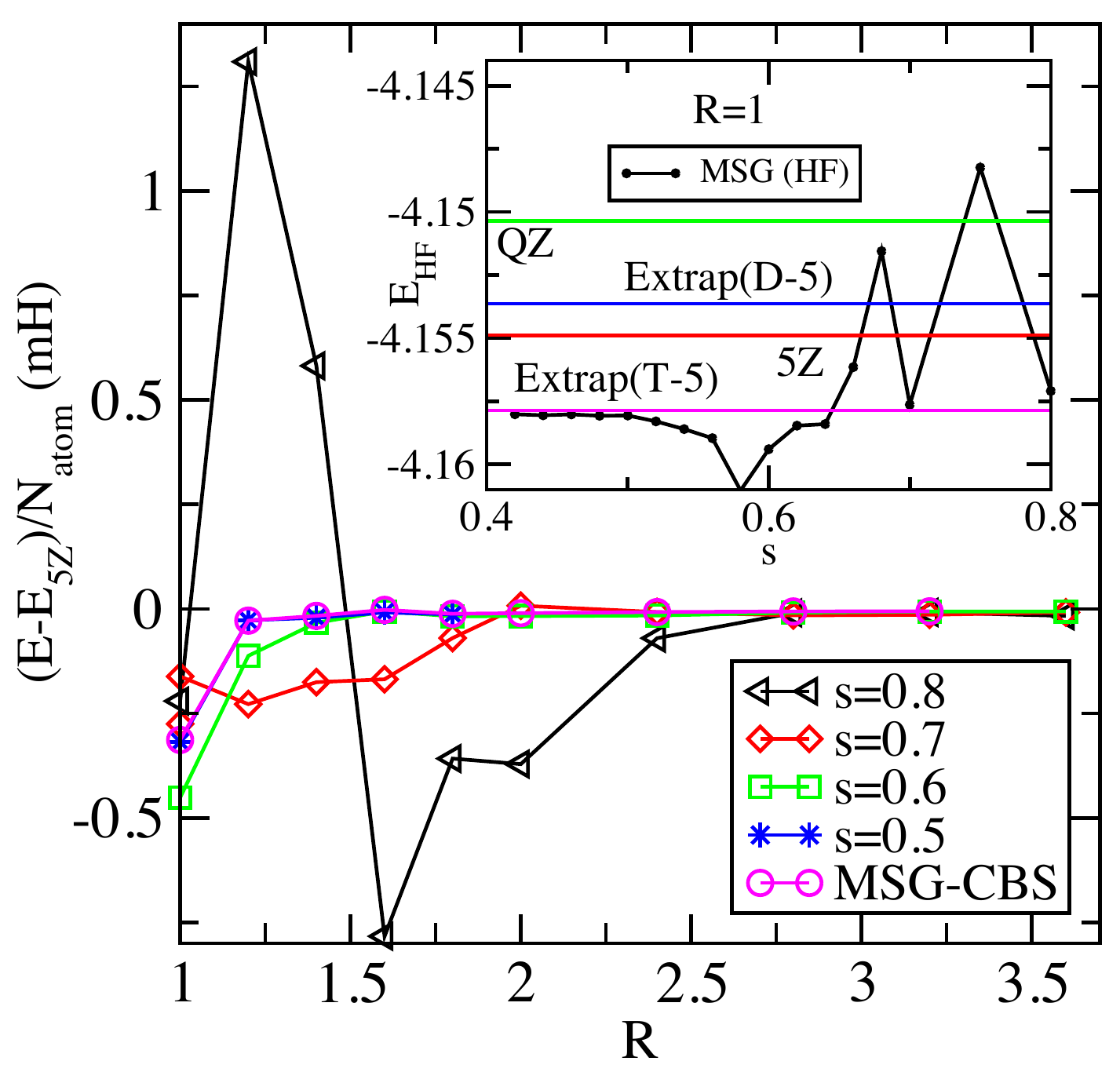}
\capt{Hartree Fock energies per atom of H$_{10}$ versus $R$, relative to the Gaussian basis
set cc-pV5Z (5Z)\cite{Motta:2017}.  The connected symbols are MSG-HF results at 
constant $s=a$, as labeled. For small $R$,
they converge to a small but noticeably different result from cc-pV5Z.  The inset shows
MSG-HF data versus $s=a$, at $R=1$. Horizontal lines show Gaussian results, along with two exponential
extrapolations, based on (TZ,QZ,5Z) (labeled T-5), and on (DZ,TZ,QZ,5Z) (labeled D-5).
The MSG-HF agrees for small $s$ with the T-5 extrapolation, although the D-5 
extrapolation was used in Ref. \cite{Motta:2017}.
} \label{HF}
\end{figure}

We now turn to a more challenging system, a linear chain of hydrogen atoms 
spaced $R$ apart.
Hydrogen chain systems were the subject of a recent benchmark study, comparing more than a dozen methods, where the main goal was to
reach the combined limit of exact correlation, complete basis set, and infinite number of 
atoms \cite{Motta:2017}.
We first consider unrestricted Hartree Fock (UHF) on H$_{10}$, shown in Fig.~\ref{HF}.
The plot shows HF energy differences relative to those of a large Gaussian 
basis, cc-pV5Z. The convergence of the MSG basis is irregular because the centers of
the gausslets are not aligned with the nuclei; but it is
easy to get very accurate results and judge the accuracy. At small $R$,
the Gaussian basis sets have trouble due to linear dependence \cite{Motta:2017},
leading to a small but noticeable discrepancy between the 5Z and MSG results.
As shown in the inset, at $R=1$ the Gaussians converge slowly, and different extrapolations
give different results.
As a rough comparison of the calculational effort for these very high accuracy
calculations: for $R=1$, $a=s=0.5$, the
MSG basis has just over $13,000$ basis functions; the number of two-electron 
integrals is the square of this, or $1.7\times10^8$. The 5Z basis has 550 functions,
but the number of integrals ($N^4$, ignoring symmetry) is $9.2\times10^{10}$. 
The calculation time of our UHF algorithm, which takes advantage of the diagonal nature
of the Hamiltonian, scales as $N^2 N_{\rm e}$, where $N_{\rm e}$ is the number of electrons,
with the dominant part coming from a Davidson diagonalization, for $N_{\rm e}$ eigenvectors, 
of the Fock matrix.

For correlated calculations,  to decrease the number of basis functions, 
one can use the HF occupied orbitals to contract the MSG basis to smaller size.
This can be done in a way that maintains the diagonal form of the interactions.  One can
also extrapolate in a cutoff that controls this contraction, to obtain results for the uncontracted
basis.  The largest systems needed for a extrapolation are still about a factor of 2 or 3 smaller
than the uncontracted basis, and the results below follow this procedure, which
 is described in the Supplementary Material.
 
We now turn to MSG-DMRG calculations for H$_{10}$.
Our DMRG implementation uses the matrix product operator compression of our earlier sliced basis DMRG (SBDMRG) 
approach \cite{Stoudenmire:2017}.
This compression makes the calculation time for fixed accuracy per atom scale linearly in the number
of atoms in a hydrogen chain both in SBDMRG and MSG-DMRG.
We are currently limited to about 3000-4000 basis functions. (In contrast, standard
DMRG in a Gaussian basis---with no diagonal approximation and no compression---is 
limited to about 100-200 active basis functions.)  We find that the DMRG performs very
well. For the very high accuracy results shown below, we generally only needed to keep about
200 states for larger $R$, and up to 400-500 for $R=1$ (due to its more metallic character). This excellent performance is due
to the high locality of the basis, which DMRG and other tensor network 
methods \cite{Verstraete:2004,Corboz:2016,ORourke:2018} strongly prefer.

\begin{figure}[t]
\includegraphics[width=0.8\columnwidth]{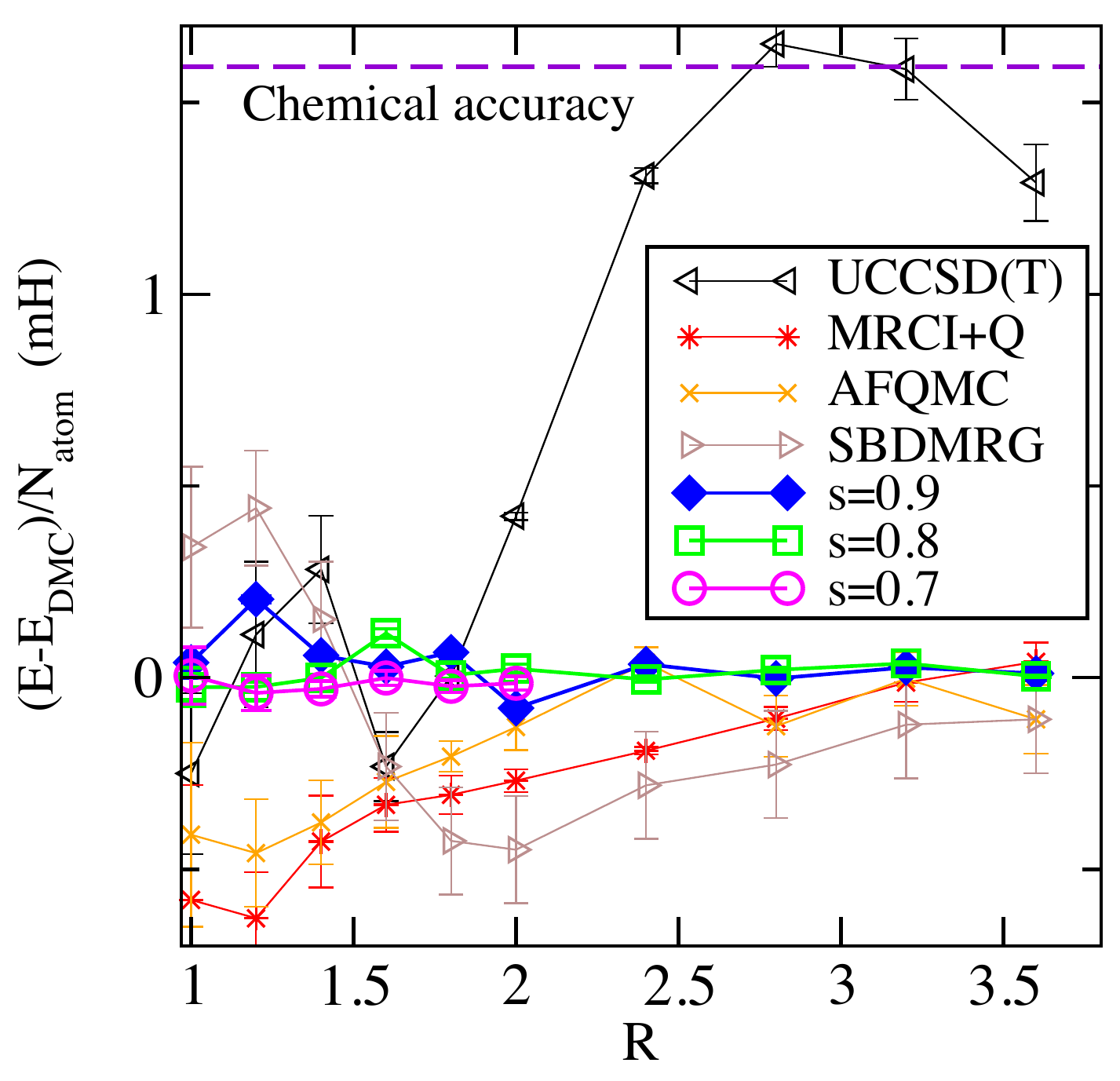}
\capt{Complete basis set energies per atom of H$_{10}$ versus $R$, relative to 
a diffusion Monte Carlo method, for MSG-DMRG (labeled by $s$) versus various approaches from Ref.~\onlinecite{Motta:2017}.   
} \label{H10}
\end{figure}

We find that the correlation energy converges faster with $s$ than the HF energy.  
This is not surprising:
the representation of the nuclear cusp is poor with a coarse gausslet basis, which is
is primarily a single particle effect.  Therefore, to get total
energies we use the HF energy with very small $s$, and add to it the correlation energy
obtained with a larger $s$, where the correlation energy is defined by subtracting the unrestricted HF energy
from the total energy for the same basis set.  

In Fig. \ref{H10}, we show a comparison of total energies for several methods \cite{Motta:2017} 
and our MSG-DMRG for various $s=a$.
All methods attempt to reach the CBS limit; for all but the MSG-DMRG and DMC methods,
this involved an extrapolation in the basis set.
The energy differences here are generally well below chemical accuracy. 
Often such high accuracy is unnecessary, but studying the high accuracy limit
is an excellent way to demonstrate the usefulness of MSG-DMRG.
The energies are measured relative to one of the diffusion Monte Carlo methods, LR-DMC-AGP (or DMC).
In Ref. \cite{Motta:2017}, at this level of accuracy, none of the best available methods
agreed, so it was not known which was best, and reference plots were made relative to 
MRCI+Q for smaller systems and AFQMC for larger ones. DMRG based on standard Gaussian
basis sets could not be done beyond the TZ level, so no CBS results were available.
Here, we find systematic convergence of MSG-DMRG to
energies agreeing with the LR-DMC-AGP method.  Agreement was poorer at small $R$ with a DMC method based
on an LDA trial function. There are systematic errors in DMC stemming from the
fixed node approximation, which are unusually small in this 1D system, but hard to quantify.
Since the nature of errors in DMC and MSG-DMRG are completely different, and since
the MSG-DMRG energies converge systematically with a control parameter, we can be
rather sure that MSG-DMRG and DMC are both getting the most accurate energies.

The MSG-DMRG errors for fixed $s=a$ are biggest at small $R$.  This is expected;
at small $R$, it would be more natural to scale $s$ with $R$, keeping the number
of basis function more nearly constant. The smallest grid
spacings are about $a\cdot s$, or about $0.5$ for $s=0.7$. Small $R$ is challenging
to the Gaussian basis set methods because the basis functions become linearly dependent.

In this strongly correlated test system in the CBS limit we have demonstrated  that
MSG-DMRG can be significantly more accurate than existing approaches based on Gaussian basis sets.  
The MSG bases can be applied to a much wider range of systems, but
more work needs to be done in going to larger-$Z$ atoms. It will also be 
interesting to explore what the favorable computational scaling of MSG bases
allows one to do when one is only interested in more limited (e.g. ``chemical'') accuracy.

We acknowledge useful conversations with Ryan Babbush, Jarrod McClean, 
Shiwei Zhang, Mario Motta, Garnet Chan, and
Sandro Sorella.
We acknowledge support from the Simons Foundation through the Many-Electron Collaboration,
and from the U.S. Department of Energy, Office of Science, Basic Energy Sciences under
award \#DE-SC008696. The Flatiron Institute is a division of the Simons Foundation.

\bibliography{msgpaper}

\end{document}


\title{Supplemental Material for 
Multi-sliced Gausslet Basis Sets for Electronic Structure}

\author{Steven R.\ White}
\affiliation{Department of Physics and Astronomy, University of California, Irvine, CA 92697-4575 USA}

\author{E.\ Miles Stoudenmire}
\affiliation{Department of Physics and Astronomy, University of California, Irvine, CA 92697-4575 USA}

\date{\today}

\maketitle

\section{Coordinate transformations for two or more atoms}

In the main text we wrote down a simple the ansatz for the local density of gausslets around a nucleus at the origin in 1D
\ben
\rho = \frac{1}{s \sqrt{x^2+a^2}} \label{oneatomdist}
\een
where $s$ sets the overall gausslet spacing and $a$ sets the core size.
It may  be desirable to limit the maximum spacing far from atoms, in which case we can modify this to
\ben
\rho = \frac{1}{s \sqrt{x^2+a^2}} +1/d \label{oneatomdistb}
\een
where $d$ is the maximum spacing.
The density at a nucleus is thus $(as)^{-1}+1/d$.
First consider a line of atoms in 1D. The $1/d$ term, if desired, should always be included only
in the final expression for $\rho$, not summed over. 
Even with the $1/d$ term omitted, if we simply add the basis function densities $\rho$ for each
atom,  we find a problem:  the slowly decaying $1/x$ tails of $\rho$ lead to too high a density in central regions.

To add densities with less buildup and to set a maximum spacing, we can consider the ansatz 
\ben
\rho(x) = \left[\sum_\alpha \rho_\alpha(x)^2\right]^{1/2}  + 1/d
\label{allrho}
\een
where $\alpha$ labels nuclei.
With this form we need to determine $u(x)$ via numerical integration.
This ansatz for $\rho$ decreases the buildup, but it still puts a greater density in the center of
a chain compared to the edges.  To fix this, in addition to using Eq. (\ref{allrho}), 
we make the core size an adjustable parameter
for each atom, $a \to a_\alpha$.  By increasing $a_\alpha$, we can decrease the function density
at nucleus $\alpha$.  We use a simple nonlinear minimization to adjust all
the $a_\alpha$ so that the density at each nucleus is a target desired density, which we take as $(as)^{-1}$, the
one atom density without any $1/d$ term.

We now consider transformations for a chain of atoms in 3D, where atoms have locations $(0,0,z_\alpha)$. 
For the $z$ transformation (there is only one) we use the recipe of the previous 
paragraph. (We would ignore the $x$ and $y$ coordinates of each atom in this case even if the
atoms were not in a chain.)
Next, consider the $y$ slicing, where we need to
choose the parameters defining $u^y_k(y)$, which determines the $y$ slicing for a specific
$z$-slice labeled by $k$.
In the case of chains, where all nuclei have the same $x$ and $y$ coordinates, 
there is effectively only one atom (at the origin) for the $x$ and $y$ slicings, and we use
Eq. (\ref{oneatomdistb}).
A slice far from all nuclei does not need the density of functions that a near-nucleus slice
does, so it should see a bigger core size, $\tilde a > a$.
Let $d_{k}$ be the distance from the nearest nucleus to $z$-slice $k$ (a plane),
i.e. the minimum of $|z_\alpha-z_k|$ over $\alpha$.
Then we define $\tilde a$ as
\ben
\tilde a=\sqrt{d_k^2+a^2},
\label{aformula}
\een
Then $u^y_k(y)$ is determined via Eq.~(\ref{oneatomdist}) with $\tilde a$ in place of $a$.
The $x$ slicing is very similar, where instead of $d_k$, we use $d_{kj}$, the distance of
the $y$ slice $(k,j)$, which is a line, to the nearest atom.

For an arbitrary 3D arrangement of atoms, the $z$ slicing would follow the same recipe.
The $y$ and $x$ slicing would need to use Eq.~(\ref{allrho}), with effective 
core sizes $\tilde a_\alpha$. For nuclei far from a slice, one would not want to
include any terms in $\rho$ to modify the density.  Precisely specifying how to do
this in an arbitrary arrangement would take some experimentation with different recipes,
which we have not done.

For almost all the calculations here we have set $d=3$. This choice was based mostly as a small multiple of the decay
length in a hydrogen atom: the wavefunction changes on a length scale of order 1, so it does not make much sense
to use any functions much larger than this scale. Changing $d$ should only affect how fast the results converge in $s$,
not what they converge to.
The exception is for H$_{10}$ at 
$R=1$.  Here, during the HF calculations, we discovered that this system has more extended
tails of the wavefunction.  This showed up when we adjusted the parameter $b$ which sets the radius around each nucleus within
which we included basis functions, finding that we needed to increase $b$ from 9 to 13 for accuracies below 0.1 mH.  
The presence of long tails suggested
that we increase $d$ also, and we ended up using $d=9$, with tests showing nearly identical HF energies for $d=7$ and $d=5$.

\section{Contractions of an MSG basis using Hartree Fock orbitals}

For correlated calculations, it is sometimes helpful or necessary to decrease the number of basis functions, 
contracting them to a smaller basis.  
The key question is:  what sorts of contractions can be performed which maintain the
diagonal form of the two particle interaction?  (It seems to be much easier to start
with a diagonal form and maintain it through contractions, 
as opposed to starting with a smaller conventional non-diagonal
basis and produce a good diagonal Hamiltonian.) Consider partitioning the MSG basis
functions into disjoint sets of functions, e.g. $\{g_1,g_2,g_4\}, \{g_3,g_5,g_{10}\}, \ldots$.
If we contract each set into a single function, e.g. specificying contraction coefficients
$\{c_1,c_2,c_3\},\ldots $, then it is simple to show that the interaction for the contracted
basis has a diagonal $V_{ij}$. Also, if one contracts to $N_b$ functions within
each set, with no contractions straddling sets, the interaction takes a block diagonal form
with $N_b^2 N'^2$ total nonzero elements, where $N'$ is the new number of functions.
This approach is reminiscint of the discontinuous Galerkin method, which has been used for density
functional theory calculations \cite{Lin:2016}; sliced basis DMRG also produce this blocked 
interaction \cite{Stoudenmire:2017}.

Given the one body reduced density matrix (RDM), also known as the 
single particle equal time Green's function, an optimal choice for the contraction coefficients
for a set would come from diagonalizing the small block of the RDM connecting the functions in
the set. The eigenvectors with maximum occupancy (i.e. eigenvalue) give the optimal contraction
coefficients.  Here, we have used the approximation for the RDM coming from the occupied orbitals
of UHF, where we spin average the RDM to get purely spatial basis functions.
This choice has the property that if the largest set occupancy which we do not keep is negligible,
then the basis contains the UHF wavefunction. The energy is at least as good as, and normally
much better than UHF.

Generally, we want to form our disjoint sets to respect spatial locality. One simple approach
is to group functions by atom, assigning each function to the closest atom. This can give
quite small basis sets, with block diagonal interactions. This approach could be very useful
for use with quantum computers \cite{Babbush:2018}.
Alternatively, we can try to maintain
the fully diagonal form with $N_b=1$, at the expense of larger $N'$.  
This is useful in trying to reach the complete basis set (CBS) limit.  
In Fig. 2(d) (main text) we show several long-thin colored blocks which would be good candidates
for contracting into one function. We can use the contractions to improve the efficiency
of the MSG basis by combining similar sets of functions, particularly in the outer tails of the 
system. Here we describe a contraction approach which reduces the $N$ by
a factor of about 2-5, which allows extrapolation in a contraction parameter, allowing us
to obtain energies of the original MSG basis, as if there were no contraction.

Let us assume we have done unrestricted Hartree Fock and gotten a set occupied spin
orbitals.  We use the occupied orbitals $\phi_{p\uparrow}(i)$, where $p$ labels orbitals
and $i$ labels MSG basis functions, to form an up reduced density matrix (RDM) as
\ben
C^\uparrow_{ij} = \sum_p \phi_{p\uparrow}(i)\phi_{p\uparrow}(j)
\label{Cup}
\een
and similarly for $C^\downarrow$.  Then we use the spin averaged RDM, which is $(C^\uparrow+C^\downarrow)/2$.

We can form a block of the RDM for any subset of the MSG basis functions, i.e.
($i_1$, $i_2$, ...). Let $w_\beta$ be the eigenvalues of this blocked RDM. Let us also
refer to this set of functions itself as a ``block". If the
block was the whole basis set, then the corresponding eigenvectors would be the natural
orbitals, forming an optimal basis (in some sense) for contraction. The eigenvectors of the block-RDM have
similar properties.  In particular, if the second eigenvalue (sorted in decreasing value; all
are non-negative) is neglible, we can contract to the first eigenvector without making
any truncation in the HF wavefunction. This contraction would replace the block
by one contracted function. The contraction coefficients would be the leading eigenvector.
If we have used the diagonal approximation for the MSG basis for the two electron interaction, 
this contraction to a single function leaves the interaction fully diagonal. This process can be
repeated for more blockings, or done simultaneously, keeping in mind that all blocks are disjoint,
i.e. no indices in common.

More practically, we can choose a cutoff $\epsilon_2$ for the size of the second eigenvalue
to control whether this contraction is allowed.  We would like to contract to as few
functions as possible subject to this cutoff for whether a contraction is allowed.  It
is most efficient to search through pairs of indices, checking whether the second eigenvalue
satisfies the cutoff.  If it does, then we can contract those two functions together.  
This can be continued, contracting pairs of functions which have themselves already been
contracted. The actual numerical transformation of the Hamiltonian matrices does not need to be done
at each step; it can be delayed to the end.  Then it is natural, and more optimal, to simply
collect the indices making up each block.  We can start by making $N$ blocks, each containing just
one index.  Then we join pairs of blocks using the cutoff criteria, but the joining is just to merge the sets
of indices.  Each diagonalization of a blocked RDM is over all the indices in the block, so that the
optimal leading eigenvector is obtained with the full freedom of the joined block.
How do we search through the pairs of blocks to find blocks to join, if they satisfy the cutoff? One natural
way is to look for nearest neighbor blocks, assuming the initial functions are sorted.  Once all near neighbor blocks
that can be paired are joined, one can look for pairings of more distantly separated blocks.  But the precise recipe for joining blocks does not appear to matter very much.

How should the MSG functions be sorted to start this procedure?  In 2D DMRG for square lattice models one typically orders sites in a way that looks like a snake: one proceeds
column by column, going up one column then down the next.  One could do the same thing in a 3D cubic lattice, in order 
to keep adjacent sites in the 1D list near each other in 3D.  One can also do the same ordering in a 3D multisliced context,
which we do.  This provides an excellent ordering for DMRG even if we do not do any contractions.

This procedure cannot recover the full natural orbitals.  Even one natural orbital would extend
over the whole system, so it would require a contraction of all functions. Instead, this procedure
tends to contract adjacent functions which are locked together, say because they are both in a
tail region of an atom and this tail could, hypothetically, be described by one extended
Gaussian basis function. Nevertheless, for a reasonable cutoff, this procedure can very substantially
reduce the size of the basis, e.g. by an order of magnitude.  In fact, it decreases the size
of the basis more than we would like, in terms of calculating the correlation energy accurately.

The $\epsilon_2$ cutoff, based on HF, ignores correlation.  If we had the exact RDM, it would not
ignore correlation and the cutoff rule would be ideal.  But, given only the HF RDM, we find that
we can get a controlled truncation which includes correlation by also using a second cutoff,
$\epsilon_1$, which depends on the first eigenvalue. The first eigenvalue is close to the
occupancy of the block, so an equally good cutoff would use the trace of the block-RDM.
Our modified procedure is to allow two functions to be joined if the first two eigenvalues are
less than $\epsilon_1$ and $\epsilon_2$, respectively.

\begin{figure}[bt]
\includegraphics[width=0.8\columnwidth]{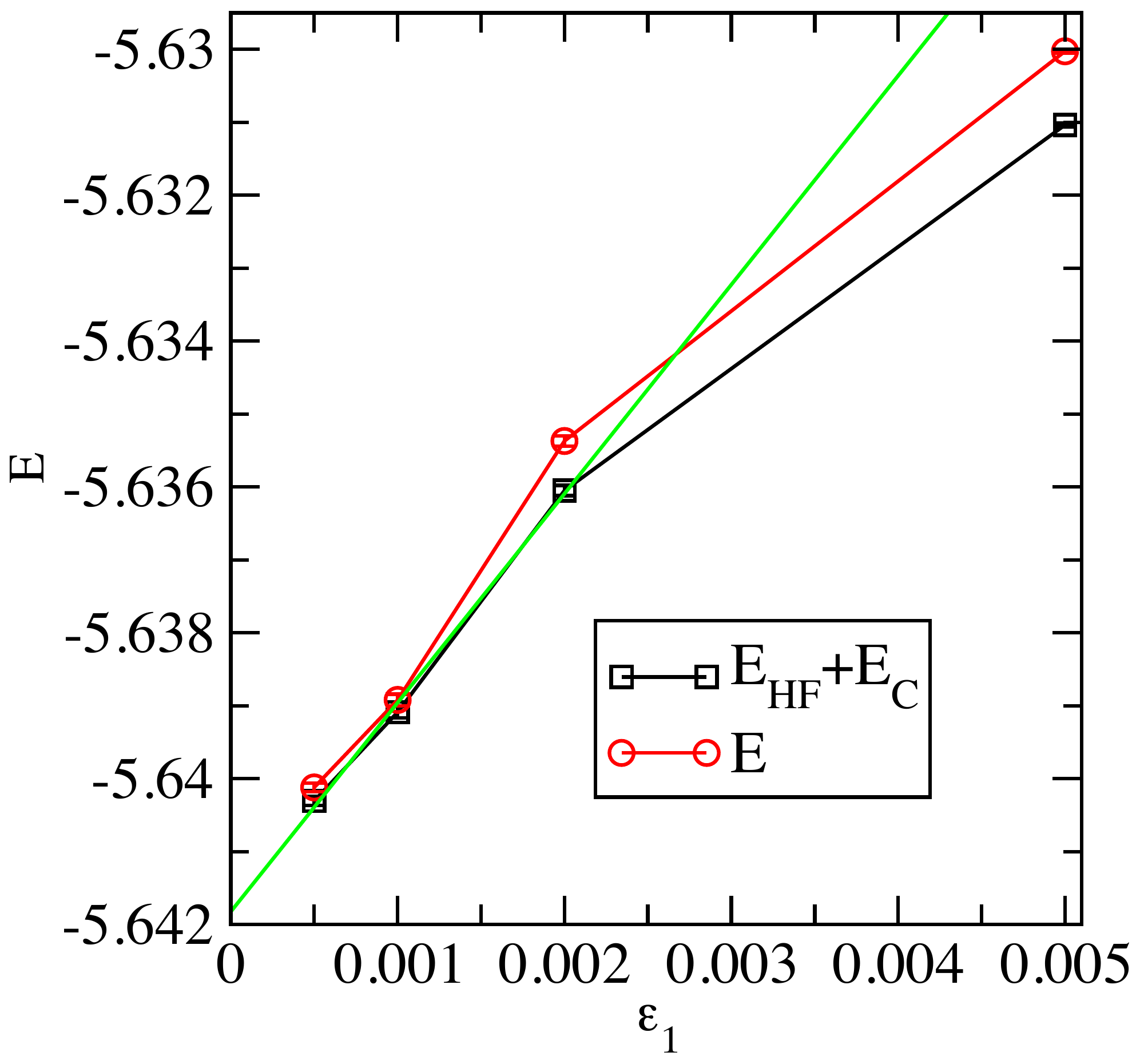}
\capt{Energy of H$_{10}$ with $R=2$, with $a=s=0.8$, versus $\epsilon_1$, where $\epsilon_2=\epsilon_1/100$.
The red circles show the energy with that cutoff, from DMRG.  The DMRG runs have been extrapolated in the
truncation error, with error bars shown.  The black squares show a separate treatment of the same energies.
Here, the correlation energy is obtained (using the same DMRG runs as for the red squares), and a separate UHF calculation
for that cutoff.  Then each correlation energy is added to the UHF energy for zero cutoff. The green line shows a linear fit
to the black points, ignoring the point with $\epsilon_1=0.005$.  This separate correlation treatment was used for the 
analysis in the main text.
} \label{Epstwo}
\end{figure}

The justification for using $\epsilon_1$ is that if the occupancy of this block is 
very small, there is very low probability for two electrons to doubly occupy the block,
so there is little need for a second function to describe the local dynamic correlation.
However, while this motivation provides a reason for trying this procedure, the key
point is that using $\epsilon_1$ and $\epsilon_2$ together works and is useful. Using
this procedure for contractions, we have explored the effects of the cutoffs, applying DMRG
so find the ground state energy of the resulting Hamiltonians. We find that
the error in the energy from these cutoffs varies linearly with the cutoff, and can be extrapolated.
The $\epsilon_2$ cutoff is more important, since it controls the representation of the HF wavefunction,
so $\epsilon_2$ should be smaller.  A reasonable choice is to lock the two together with a fixed ratio.
In  Fig. \ref{Epstwo} we show results for a typical $H_{10}$ system with $\epsilon_2=\epsilon_1/100$.
For small enough cutoffs, we see linear dependence of the energy  on the cutoff, which can allow reasonable
extrapolation, as shown. 

The extrapolations are somewhat improved by separating out the correlation energy and extrapolating that,
rather than the total energy.  The correlation energy can be added to the HF energy without any contractions
for the MSG basis. This gives the energy for a particular MSG basis, primarily defined by the parameters 
$a$ and $s$, where we normally take $a=s$.   As discussed in the main text, one can then also separate out the correlation energy to 
improve the convergence with $a=s$.  In this case, we take the correlation energy for say $a=s=0.8$, and
add it to the HF energy obtained from much smaller $a=s$.  This procedure was used for the $H_{10}$ results in the main text.

\section{Evaluation of Hamiltonian Matrix elements for the MSG basis}

The key ``trick'' in evaluating matrix elements is to replace the Coulomb interaction by a sum of Gaussians, a trick
which has been known for a long time. Specifically, one can write
\ben
1/r = \sum_i c_i e^{-a_i r^2}
\label{gaucoulomb}
\een
with properly chosen coefficients $a_i$ and $c_i$.  Since the integral evaluation ends up being a minor part of
our computations, we have chosen a very accurate 220 term representation that is very accurate (better than $10^{-8}$)
over a wide range of $r$ ($10^{-8}$ to $10^4$). Representations with fewer terms are known and would be useful if
the matrix element evaluation became too time-consuming.

This trick separates 3D integrals into products of 1D integrals for basis functions which are in the coordinate
product form, as MSG functions are.  For example,
\bea
\int_{\vec r} e^{-a_i r^2} f(x) g(y) h(z) =\qquad\qquad\qquad\qquad\\
\int dx\  e^{-a_i x^2} f(x)
\int dy\  e^{-a_i y^2} g(y)
\int dz\  e^{-a_i z^2} h(z) \nonumber .
\label{gauseparate}
\eea
One performs the sum over $i$ as the outermost operation.

For the two electron interaction the trick replaces 6D integrals by 2D integrals.  For example,
\bea
\int_{\vec r,\vec r'} e^{-a_i (\vec r-\vec r')^2} f_1(x) g_1(y) h_1(z) f_2(x') g_2(y') h_2(z')\nonumber  \\
=
\left [\int dx dx' e^{-a_i (x-x')^2} f_1(x)f_2(x')\right ] \ldots\qquad\qquad
\label{gauseptwo}
\eea
where the dots represent similar terms for $y$ and $z$.

There are many such 2D integrals to do, so it is important to do these efficiently.
There are a wide variety of approaches one could take for this computational task.
Our approach was loosely based on the definition of 1D gausslets as a sum of
equal-width Gaussians centered on a uniform grid. This defining array-of-Gaussians
representation is an excellent basis in many respects, except that it is not orthogonal.  But for evaluating
integrals, nonorthogonality of an underlying representation is not important.
The widths of the Gaussians is chosen to be the grid spacing, but any fixed width bigger than about this would
make an excellent basis able to represent almost any function smooth at the scale of the grid.
If there were no coordinate
transformations, this representation would immediately translate integrals over
the 1D basis functions into sums of analytic integrals of the underlying Gaussians.

Unfortunately, the coordinate transformations distort the representation into a nonuniform
sum over distorted Gaussians, for which there are no analytic integrals.  So, instead,
we create a representation of the basis functions as a sum of undistorted 
but unequal width, unequally spaced Gaussians. The widths and locations of come from
a coordinate transformation.

We first create coordinate transformations that put a higher density of functions near nuclei.
Instead of multislicing, we choose exactly three 1D coordinate transformations, for $x$, $y$, and $z$.
These coordinate transformations are choosen so that the local spacing between grid points is sufficiently
small to represent all details of the 1D basis functions arising in the multislicing. This  makes these
grids finer than the basis grid where the MSG functions live by at least an order of magnitude.  For the case
of the $x$ direction, we have a function $x(u)$, with grid points $x_i = x(i)$, and local spacing $dx/du$. 
We put a Gaussian at each grid point, with a width given by $\alpha dx/du$. We choose $\alpha > 1$ to make up
for the unequal grid spacings; for a uniform grid, $\alpha=1$ is fine.
We find that $\alpha=1.25$ is a good choice, and we find that it is possible to rerepresent all our 1D basis functions
in terms of this array-of-Gaussians representation with high accuracy. We find the coefficients of the
Gaussians for each 1D basis function using a least squares procedure. The unequal spacing means that there are
many fewer degrees of freedom than the simplest approach one might take, using  a uniform very fine grid to represent
all the 1D functions.

Once all functions are defined in terms of Gaussians, all integrals (kinetic, and one and two particle potential integrals) 
are analytic, and the main work is summing over the various terms.  We implemented the Hamiltonian
construction in the Julia language,\cite{Julia} 
and our code runs fast enough to take much less time than the DMRG
calculations and also  usually less time than the HF calculations.

\bibliography{supp}